\documentstyle[masaps]{revtex}

\newcommand{\PSbox}[3]{\mbox{\rule{0in}{#3}\includegraphics{#1}\hspace{#2}}}

\title{New physics of grain boundaries in bcc metals from the atomic-level:
molybdenum as a case study}

\author{D. Ye\c{s}illeten\dag \and T.A. Arias\ddag}

\address{\dag Departments of Physics, Massachusetts Institute of Technology, 
Cambridge, MA 02139\\
\ddag Laboratory of Atomic and Solid State Physics, Cornell University, 
Ithaca, NY 14853\\
\ddag Research Laboratory of Electronics, Massachusetts Institute of Technology, Cambridge, MA 02139}

\date{\today}

\abstract{We present a systematic trend study of the symmetric tilt
grain boundaries about the $\langle 110 \rangle$ axis in molybdenum.
Our results show that multiple structural phases, some incorporating
vacancies, compete for the boundary ground state.  We find that at low
external stress vacancies prefer to bind to the boundaries in high
concentrations, and moreover, that external stress drives structural
phase transitions which correspond to switching the boundaries on and
off as pipe-diffusion pathways for vacancies.  Finally, we present
physical arguments which indicate these phenomena are likely to occur
in the other bcc transition metals as well.}

\twocolumn

\begin{document}

\maketitle

\section{Introduction}

Molybdenum, with its high melting point and relatively inert chemical
nature is often considered for high-temperature structural
applications, but its extreme brittleness limits its usefulness.
Experimental studies\cite{kkky,koky} demonstrate that this brittleness
is an intrinsic property of the material and largely unrelated to the
presence of impurities.  These studies suggest, moreover, that this
brittleness arises from weak inter-granular cohesion along the grain
boundaries\cite{tty}.  In this work, we shed light on the microscopic
physics of these boundaries by presenting a detailed, atomic-level
trend study of the behavior and structure of their low energy phases
and the transitions among these phases.

This study reveals new physics in the interaction of the grain
boundaries with vacancies.  The traditional mechanisms of interplay
between vacancies and grain boundaries include pipe diffusion of
vacancies along the boundary\cite{mic,ic1,ic2}, and absorption and
emission of vacancies during continuous climb of primary or secondary
dislocations\cite{ball1,ball2,hirth,sutton1}.  We find, in addition,
that boundary vacancies prefer to collect together at high densities
on the boundary plane.  Our results also indicate that grain
boundaries can either emit or absorb large concentrations of vacancies
into the surrounding bulk while undergoing structural phase
transitions under applied stress.  While our focus in the present
study is on the particular system of symmetric tilt boundaries around
the $\langle 110 \rangle$ axis in molybdenum, which are known to
dominate the recrystallization texture of this material\cite{tty}, we
expect for reasons detailed below that these conclusions hold quite
generally for symmetric tilt boundaries in bcc materials.

\section{Procedure} \label{sec:proc}

The heavy computational demands of full-blown {\em ab initio}
electronic structure calculations\cite{payne} and semi-empirical
tight-binding models\cite{carlsson,petti,foiles,xuadams} restrict
their use to the study of relatively small systems and relatively few
configurations.  In order to understand complex processes such as
fracture, dislocation migration and inter-granular cohesion,
computationally more feasible empirical potentials must be used.
Moriarty has developed such an empirical model based on a multi-ion
interatomic potential developed from first principles generalized
pseudopotential theory\cite{mori1}.  The resulting model generalized
pseudopotential theory (MGPT) potential successfully predicts the
cohesive, structural, elastic, vibrational, thermal and melting
properties of molybdenum\cite{mori3}, as well as the ideal shear
strength and self-interstitial and vacancy formation
energies\cite{mori4}.  We use this potential throughout this work.

Focusing on the $\langle 110 \rangle$ symmetric tilt boundaries, we
consider $\Sigma$=3(112) and $\Sigma$=9(114), which are among the
lowest in energy, and $\Sigma$=3(111), $\Sigma$=9(221),
$\Sigma$=11(113) and $\Sigma$=11(332) as examples of boundaries with
higher energies.  To study the physics of these boundaries, which
reside in bulk material, we employ periodic boundary conditions as the
most natural.  To minimize boundary-image interactions, we always
maintain at least seventeen layers of atoms between boundaries in our
supercells.

Determination of the ground state and the low energy excited state
structures in principle requires the exploration of the phase space of
all possible configurations, which is an impractical task without
taking into account some basic physics.  The primary consideration we
use to restrict this phase space is that, due to the relatively strong
directional bonding in molybdenum and similar bcc metals, the
structure of the grain boundaries tends to preserve the {\em internal}
topology of individual grains.  Under this restriction, there remain
then only three considerations for each boundary: (1) possible
addition and removal of atoms to and from the faces of the grains at
the boundary, (2) the displacement of the grains relative to one
another, and (3) relaxation of the internal atomic coordinates.

\begin{figure} 
\begin{center}
\PSbox{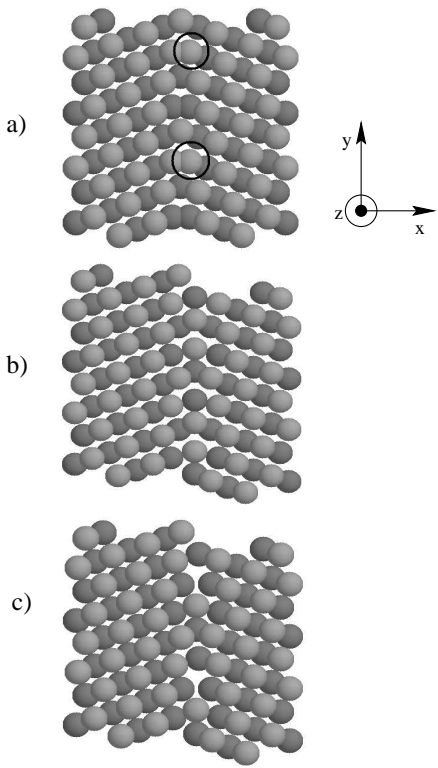 vscale=45 hscale=45 hoffset=40 voffset=0}{3in}{4in}
\end{center}
\caption{Structural phases of the $\Sigma$=9(114) grain boundary: (a)
na\"{\i}ve CSL boundary, (b) relaxed full-material boundary, (c)
relaxed vacancy phase boundary.  (To aid visualization, atoms from the
two cubic sublattices are colored separately, light and dark.)}
\label{sig9}
\end{figure}

For the first consideration, the fact that the interstitial energy in
molybdenum ($\geq$ 10eV) is much larger than the vacancy energy
(3eV)\cite{mori4} indicates that insertion of additional material at
the boundary leads to unlikely high interfacial energies.  We
therefore concentrate only on the {\em removal} of atoms at the
boundary.  Direct calculations with the MGPT potential reveal that the
most favorable sites for atom removal are in the vicinity of the
boundary.  It turns out that, because the atoms in the planes adjacent
to the boundary plane (indicated by circles in Figure~\ref{sig9}a)
pack closely together, these sites are the energetically most
favorable for vacancy formation.

This leads us to consider the following structural phases for the
boundaries in our study: grains joined with the amount of material
expected from the {\em na\"{\i}ve} coincident site lattice (CSL)
construction (``{\em full-material phase}''), and boundaries where we
remove atoms from the circled sites in Figure~\ref{sig9}a, which shows
this construction.  Below, we find that binding energy per boundary
vacancy is higher when vacancies collect together at high density on
the boundaries.  Therefore, we first concentrate on boundaries with
high vacancy densities.  Because removal of an entire plane of atoms
near the boundary is topologically equivalent to the initial ``full
material'' phase under appropriate relative displacement of the
grains, we focus on the phase where we remove one half-plane of atoms
from the layer adjacent to the boundary (``{\em vacancy phase}'').

To determine the ground state of the above two structural phases, we
next turn to the second consideration above, the relative shifts of
the grains.  The displacement-shift complete (DSC) cell, which we
explore on a $16 \times 16$ sampling grid (which reduces to $4 \times
4$ by symmetry), contains all possible unique relative planar shifts
of the grains.  For each such shift, we complete our search and
address the third and final consideration by performing full
relaxations of both the perpendicular expansion of the grains and
their internal coordinates.  This extensive survey requires force and
energy calculations of approximately 100,000 boundary configurations,
and would be infeasible to carry out with electronic structure
techniques.

To confirm the effectiveness of this survey in identifying ground
state structures, we repeat the above procedure with supercells in
which we remove an entire plane of atoms from the boundary.  For all
six boundaries in our study, our procedure indeed identifies the
appropriate shift to recover the initial, topologically equivalent
ground state found for the full-material phase before the removal of
the plane of atoms.

\section{Results}

\subsection{Low Energy Phases}

Our results reveal several general trends in the physics of the
$\langle 110 \rangle$ tilt grain boundaries in molybdenum.  As a
specific example, consider the $\Sigma$=9(114) boundary, which
Figure~\ref{sig9} shows.  Panel (a) shows the na\"{\i}ve CSL
construction of the full-material phase, whose ground state as
identified through our procedure, appears in Figure~\ref{sig9}b.  We
find that this phase lowers its energy through both a perpendicular
expansion of the boundary and integranular shifts parallel to the
boundary, both of which tend to increase the local volume for the
closely packed atoms near the boundary plane, restoring them to a more
bulk-like environment.

Table~\ref{tbl:shift} shows that the outward expansion is a general
trend among all grain boundaries in our study and that shifts occur
along the boundary in the direction perpendicular to the tilt axis
($y-$direction, Figure~\protect{\ref{sig9}}) to allow for a more
bulk-like local environment.  We find no significant shifts along the
tilt axis ($z-$direction, Figure~\protect{\ref{sig9}}) for this phase
of any of the boundaries.  Finally, the last column of the table gives
the mechanical compliance ($(1/k)$) of each boundary, where $k$ is
determined from the quadratic response of the energy per unit area per
boundary as the cell expands, which takes the form
\begin{equation}
\Delta{\cal U}=\Delta{\cal U}_o+\frac{1}{2}k(x-\Delta x)^2+\ldots
\end{equation}
where $x$ is the expansion of the cell, and $\Delta x$ and
$\Delta{\cal U}_o$ are the relaxed perpendicular expansion and
boundary energy given in the table, respectively.  The table lists the
difference between $1/k$ for the boundary and for the same cell filled
with bulk, as it is this difference which defines the response of the
boundary, {\em independent} of the bulk content of the cell.

Turning now to the vacancy phase, Figure~\ref{sig9}c shows the results
of our ground-state search for the $\Sigma$9(114) boundary as formed
by removing the circled atoms from Figure~\ref{sig9}a.  We now find
grain-shifting as well as local internal relaxations to create more
bulk-like local atomic environments, and as Table~\ref{tbl:vac}
summarizes, we again find this behavior for all boundaries in our
study.  The perpendicular shift is always {\em inward} compared to the
ground state of the full-material phase, so as to close the material
void associated with the vacancies.  Finally, we frequently find for
the vacancy phase, even for the highly stable $\Sigma$3's, parallel
shifts relative to the full-material phase which produce more natural
bonding arrangements for the boundary to accommodate the vacancies.
This tendency for accommodation is so strong as to induce the only
shift along the tilt axis we observe in this study, for the vacancy
phase of the $\Sigma11(332)$ boundary.

\begin{figure} 
\begin{center}
\PSbox{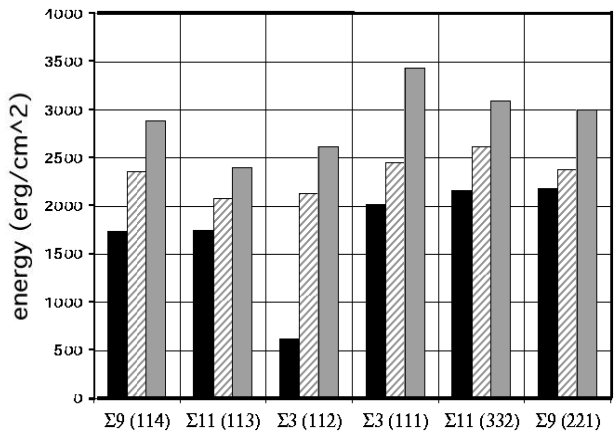 vscale=75 hscale=70 hoffset=-110 voffset=-210}{3in}{2.4in}
\end{center}
\caption{Energetics of grain boundary phases: full-material phase
(black bars), vacancy phase (hatched bars), bulk-vacancy phase (gray
bars).}
\label{compen}
\end{figure}

Figure~\ref{compen} compares the ground-state boundary energies of the
various phases.  In all cases, we find the full-material phase (black
bars) to be lower in energy than the vacancy phase (hatched bars),
although often not by far.  As in the experimental case, the lowest
ground state interfacial energy occurs for the naturally occurring
twin $\Sigma$3 (112) boundary\cite{tty}.  Moreover, apart from this
twin, all remaining ground state boundary energies are fairly constant
(within 25\%), as also found in experiment (within 30\%\cite{tty}).

Figure~\ref{compen} presents another relevant comparison.  To
transform physically into the full-material phase, the vacancy phase
must first expel its vacancies into the surrounding bulk material.
Thus, when considering transitions, the relevant comparison is between
the vacancy phase and a third phase, which consists of the
full-material phase plus the corresponding number of vacancies in the
surrounding bulk material ({\em bulk-vacancy phase}).  (Table
\ref{tbl:isovac} provides the relevant information for the resulting
bulk vacancies of this third phase.)  Figure~\ref{compen} shows that,
although creation of vacancies on the boundary always increases the
boundary energy, the energy for creating the corresponding number of
vacancies in the bulk is always higher.  Our results therefore are
consistent with the fact that the boundaries act as reservoirs for
vacancies, as occurs during pipe diffusion.

To verify, as mentioned above in Section II, that boundary vacancies
indeed prefer to cluster together, we have also considered boundaries
with nearly isolated vacancies within our supercell approach.
Table~\ref{tbl:singvac} presents energy, displacement and compliance
results for boundaries where the vacancy concentration (3.3\%) is one
fifteenth that of the boundary-vacancy phase.  Table~\ref{tbl:bind}
presents the analysis of this data.  As the larger binding energies
reflect, the concentrated phase is more stable.  This added stability
appears to arise from the structural relaxation through parallel
shifting of the grains observed in Table~\ref{tbl:vac}, made possible
by the high density of vacancies.  Finally, we note that in the
extreme case of the $\Sigma 11$ boundaries, vacancies do not even bind
to the boundary at low densities.

\subsection{Phase Transitions}

Three trends which the preceding results exhibit can be expected from
general, material-independent considerations: (1) that the boundaries
present preferred binding sites for vacancies because they disrupt the
bulk bond-order, (2) that boundary vacancies prefer to collect into
the high-density vacancy phases because of the additional relaxation,
which parallel shifting of the grains affords, and (3) that the
binding of vacancies to the boundary reduces the intergranular spacing
because this restores more bulk-like interatomic separations.  These
phenomena open the intriguing possibility that the application of
tensile stress normal to grain boundaries in bcc metals generally
drives transitions among these various structural phases, thereby
providing new forms of boundary-vacancy interaction.

To explore this possibility, we consider the thermodynamic potential
which is minimized under fixed external stress, the enthalpy.  As a
function of applied perpendicular stress $\sigma$, the enthalpy of a
grain boundary structure relative to bulk is
\begin{equation}
\Delta{\cal H} = \Delta {\cal U}_o - \sigma \, \Delta x -
\frac{\sigma^2}{2} \Delta (1/k) + {\mathcal O}(\sigma^3). 
\end{equation}
Here $\Delta {\cal U}_o$ is the difference in the ground-state energy
per unit area, $\Delta x$ is the difference in preferred perpendicular
intergranular separation, and $\Delta (1/k)$ is the difference in
compliance.  Each of these materials parameters appears for each
relevant phase in Tables \ref{tbl:shift}-\ref{tbl:singvac}.

\begin{figure} 
\begin{center}
\PSbox{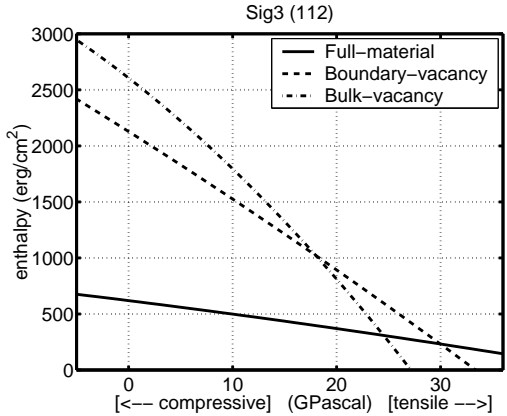 vscale=36 hscale=36 hoffset=-20 voffset=-60}{2.8in}{2.1in}
\caption{Enthalpies of $\Sigma3(112)$ boundary as a function external
stress for all three phases.}
\label{gbenths}
\end{center}
\end{figure}

As an example, Figure~\ref{gbenths} shows the behavior of the
enthalpy, relative to bulk, of the naturally occurring $\Sigma3(112)$
boundary in its full-material, vacancy and bulk-vacancy phases.  Three
first-order phase transitions (enthalpy crossings) are evident in the
figure.  At zero stress, as observed above, the full-material phase is
the ground state of the boundary, and moreover, in the presence of
vacancies, the vacancy phase has lower energy than the bulk-vacancy
phase, indicating that vacancies prefer the boundary over the bulk.
However, at an applied stress of about 18~GPa, the $\Sigma3(112)$
boundary system undergoes a first-order phase transition in which the
vacancy phase is no longer preferred, and the boundary ejects its
vacancies into the surrounding bulk material.  A second transition
occurs near 25~GPa, at which point the bulk-vacancy phase becomes
lower in enthalpy than the bulk material phase.  This corresponds to
the {\em spontaneous} formation of vacancies in bulk, indicating
breakdown of the bulk material.  The third transition occurs near
30~GPa.  Were this transition accessible before the breakdown of the
bulk material, it would correspond to spontaneous formation of
boundary vacancies.

Table~\ref{tbl:trans} presents the stresses for the above three
transitions for all boundaries in our study.  In all cases, the
emission stress is accessible before breakdown of the bulk material
through spontaneous formation of vacancies.  Moreover, except the for
the outlying behavior of the $\Sigma9(221)$ boundary, vacancies always
first form spontaneously in the bulk before they do so on the
boundaries.

To verify that these results are not artifacts of the high-density
vacancy phase, we repeat the above enthalpy analysis for boundaries
with low densities of vacancies using the data of
Table~\ref{tbl:singvac}.  As Table \ref{tbl:transsing} summarizes, we
again observe the same transitions.  The $\Sigma11$ boundaries do not
bind vacancies at low concentrations (Table~\ref{tbl:bind}), and
therefore, the transition stresses for the emission of vacancies for
these boundaries are not physically relevant.  For the lower $\Sigma$
boundaries (except the $\Sigma9(221)$ boundary, which again exhibits
an outlying behavior), diluting the vacancy concentration reduces the
critical emission stress at which the boundaries emit boundary
vacancies into the bulk ($\sigma_c^{emit}$), thus making this
transition more accessible.

Finally, as we expect from the fact that boundary vacancies are more
stable in high concentrations, the stress required to induce formation
of vacancies on the boundaries ($\sigma_c^{tear}$) at low densities is
generally greater than for the high-density vacancy phase.  As a last
consistency check on our analysis, we note that the breakdown stress
for the bulk ($\sigma_c^{break}$), as a characteristic of the perfect
crystal and not the boundary, remains essentially unchanged between
the two independent sets of calculations for low and high boundary
vacancy concentrations.

\section{Conclusions}

Several general conclusions reached in the discussion above are
supported both by trends in our calculations for a specific set of
boundaries in molybdenum and by quite general, material independent
arguments.  We expect that the following conclusions likely hold
generally for the interactions among vacancies and tilt grain
boundaries in bcc transition metals: (1) consistent with the
traditional view of grain boundaries as diffusion pathways, vacancies
prefer the boundaries over the bulk at low stresses, (2) boundary
vacancies prefer to collect into high-density vacancy phases, (3)
application of sufficient tensile stress to a boundary induces a
structural phase transition which drives the vacancies from the
boundaries into the bulk, thereby shutting off pipe diffusion along
the boundary.  The last of these conclusions in particular may have
important implications for crack growth through pipe-diffusion
assisted void growth and void formation at grain boundaries.

Finally, in terms of precise quantitative predictions of the critical
stresses characterizing these phenomena, it is important to bear in
mind that the particular interatomic potential which we have employed
(MGPT), although one of the most reliable, is known to exaggerate
energy scales for complex structures\cite{sohrab}.  We therefore would
expect to find these same transitions, but most likely at lower
stresses, when studied either experimentally or {\em ab initio}.

\acknowledgements
This work was supported by an ASCI ASAP Level 2 grant (contract
\#B338297 and \#B347887).  We thank members of the
H-division at Lawrence Livermore National Laboratories for providing
the Mo MGPT code and for many useful discussions.

\begin{table} 
\begin{center}
\begin{tabular}{c||c|c|c|c|c}
Boundary & $\Delta {\cal U}_o$ & $\Delta x$ & $\Delta y$ & $\Delta z$ & $\Delta (1/k)$  \\
& [mJ/m$^2$] & [\AA] & [\AA] & [\AA]  & $[$mJ/m$^2$/Pa$^2]$  \\ \hline \hline
$\Sigma3(112)$ &  610 &  0.1 & 0.3  & 0.0 &  900  \\ \hline
$\Sigma3(111)$ & 2020 &  0.5 & 0.3  & 0.0 & 4770  \\ \hline
$\Sigma9(114)$ & 1730 &  0.5 & 0.1  & 0.0 & 3410  \\ \hline
$\Sigma9(221)$ & 2180 &  0.4 & 0.1  & 0.0 &  100  \\ \hline 
$\Sigma11(113)$& 1740 &  0.5 & 0.2  & 0.0 & 1460  \\ \hline
$\Sigma11(332)$& 2160 &  0.4 & 0.5  & 0.0 &  580  \\
\end{tabular}
\caption{Full-material phase: energies ($\Delta {\cal U}_o$),
perpendicular expansions ($\Delta x$), and shifts ($\Delta y$ and
$\Delta z$) relative to the CSL construction, and compliances relative
to bulk ($\Delta (1/k)$), where coordinates are as defined in
Figure~\protect{\ref{sig9}}.}
\label{tbl:shift}
\end{center}
\end{table}

\begin{table} 
\begin{center}
\begin{tabular}{c||c|c|c|c|c}
Boundary & $\Delta {\cal U}_o$ & $\Delta x$ & $\Delta y$ & $\Delta z$ & $\Delta
(1/k)$ \\ & [mJ/m$^2$] & [\AA] & [\AA] & [\AA] & $[$mJ/m$^2$/Pa$^2]$
\\ \hline \hline 
$\Sigma3(112)$ & 2130 & -0.1 & 0.4 & 0.0 & 3030 \\ \hline 
$\Sigma3(111)$ & 2450 & 0.1 & 0.3 & 0.0 & 1440 \\ \hline
$\Sigma9(114)$ & 2360 & 0.3 & 0.1 & 0.0 & 7250 \\ \hline
$\Sigma9(221)$ & 2380 & 0.3 & 0.9 & 0.0 & 2210 \\ \hline
$\Sigma11(113)$& 2070 & 0.3 & 0.8 & 0.0 & 2310 \\ \hline
$\Sigma11(332)$& 2610 & 0.1 & 0.4 & 0.8 & 1910 \\
\end{tabular}
\caption{Vacancy phase: energies ($\Delta {\cal U}_o$), perpendicular
expansions ($\Delta x$), and shifts ($\Delta y$ and $\Delta z$)
relative to the CSL construction, and compliances relative to bulk
($\Delta (1/k)$), where coordinates are as defined in
Figure~\protect{\ref{sig9}}.}
\label{tbl:vac}
\end{center}
\end{table}

\begin{table} 
\begin{center}
\begin{tabular}{c||c|c|c}
Boundary & $\Delta {\cal U}_o$ & $\Delta x$ & $\Delta (1/k)$  \\
& [mJ/m$^2$] & [\AA] & $[$mJ/m$^2$/Pa$^2]$  \\ \hline \hline
$\Sigma3(112)$ & 130 & -0.003 & 1050  \\ \hline
$\Sigma3(111)$ & 100 & -0.002 &  870  \\ \hline
$\Sigma9(114)$ &  80 &  0.003 &  500  \\ \hline
$\Sigma9(221)$ &  60 & -0.001 &  470  \\ \hline 
$\Sigma11(113)$&  50 & -0.001 &  130  \\ \hline
$\Sigma11(332)$&  60 & -0.005 &  180  \\
\end{tabular}
\caption{Isolated vacancy enthalpy information for each boundary
orientation.  Results are expressed for the number of
vacancies per unit area of the corresponding boundary-vacancy phase.
Lattice expansion $\Delta x$ and compliance $\Delta (1/k)$ information
is for longitudinal strain perpendicular to the boundary plane.}
\label{tbl:isovac}
\end{center}
\end{table}

\begin{table} 
\begin{center}
\begin{tabular}{c||c|c|c|c|c}
Boundary & $\Delta {\cal U}_o$ & $\Delta x$ & $\Delta y$ & $\Delta z$ & $\Delta
(1/k)$ \\ 
& [mJ/m$^2$] & [\AA] & [\AA] & [\AA] & $[$mJ/m$^2$/Pa$^2]$
\\ \hline \hline 
$\Sigma3(112)$ &  720 & 0.1 & 0.3 & 0.0 &  950 \\ \hline 
$\Sigma3(111)$ & 2076 & 0.5 & 0.3 & 0.0 & 2060 \\ \hline
$\Sigma9(114)$ & 2360 & 0.4 & 0.1 & 0.0 & 3520 \\ \hline
$\Sigma9(221)$ & 2200 & 0.3 & 0.1 & 0.0 &  260 \\ \hline
$\Sigma11(113)$& 1900 & 0.5 & 0.2 & 0.0 & 1900 \\ \hline
$\Sigma11(332)$& 2250 & 0.4 & 0.5 & 0.0 &  990 \\
\end{tabular}
\caption{Bulk-vacancy phase: energies ($\Delta {\cal U}_o$),
perpendicular expansions ($\Delta x$), and shifts ($\Delta y$ and
$\Delta z$) relative to the CSL construction, and compliances relative
to bulk ($\Delta (1/k)$), where coordinates are as defined in
Figure~\protect{\ref{sig9}}.}
\label{tbl:singvac}
\end{center}
\end{table}

\begin{table} 
\begin{center}
\begin{tabular}{c||c|c}
Boundary & Binding energy & Binding energy \\
         & per vacancy at & per vacancy at \\
         & high density   & low density \\ 
& [eV] & [eV] \\ \hline \hline 
$\Sigma3(112)$ & 0.7 &   0.6 \\ \hline
$\Sigma3(111)$ & 2.1 &   1.1 \\ \hline
$\Sigma9(114)$ & 1.4 &   0.7 \\ \hline
$\Sigma9(221)$ & 2.3 &   2.0 \\ \hline
$\Sigma11(113)$& 1.4 &  -7.1 \\ \hline 
$\Sigma11(332)$& 1.4 &  -1.7 \\ 
\end{tabular}
\caption{Boundary-vacancy binding energies at high and low
densities. (The $\Sigma11$ boundaries do not bind vacancies at low
concentrations.)}
\label{tbl:bind}
\end{center}
\end{table}

\begin{table}
\begin{center}
\begin{tabular}{c||c|c|c}
Boundary & $\sigma_c^{emit}$ & $\sigma_c^{break}$ & $\sigma_c^{tear}$ \\
& $[$GPa$]$ & $[$GPa$]$ & $[$GPa$]$   \\ \hline \hline
$\Sigma3(112)$&   18   & 25   &  30   \\ \hline
$\Sigma3(111)$&   17   & 24   &  not observed \\ \hline
$\Sigma9(114)$&   21   & 23   &  25   \\ \hline
$\Sigma9(221)$&   38   & 24   &  10   \\ \hline
$\Sigma11(113)$&  21   & 28   &  42   \\ \hline
$\Sigma11(332)$&  31   & 33   &  35   \\
\end{tabular}
\caption{Critical stresses for the the phase transitions discussed
in the text: emission of vacancies from the boundary into the bulk
($\sigma_c^{emit}$), breakdown of the bulk through spontaneous
formation of vacancies ($\sigma_c^{break}$), formation of vacancies,
and thus tearing, at the boundary ($\sigma_c^{tear}$).}
\label{tbl:trans}
\end{center}
\end{table}

\begin{table}
\begin{center}
\begin{tabular}{c||c|c|c}
Boundary & $\sigma_c^{emit}$ & $\sigma_c^{break}$ & $\sigma_c^{tear}$ \\
& $[$GPa$]$ & $[$GPa$]$ & $[$GPa$]$   \\ \hline \hline
$\Sigma3(112)$&   16   & 25   &  32   \\ \hline
$\Sigma3(111)$&   12   & 24   &  not observed \\ \hline
$\Sigma9(114)$&    9   & 23   &  58   \\ \hline
$\Sigma9(221)$&   44   & 24   &   8   \\ \hline
$\Sigma11(113)$&  90   & 28   &  60   \\ \hline
$\Sigma11(332)$&  125  & 33   &  52   \\
\end{tabular}
\caption{Critical stresses for the same transitions as in
Table~\ref{tbl:trans}, but at 3.3\% vacancy concentration.}
\label{tbl:transsing}
\end{center}
\end{table}

\end{document}